# Linear Software Models: Key Ideas

## Iaakov Exman

Software Engineering Department
The Jerusalem College of Engineering – JCE – Azrieli
POB 3566, Jerusalem, 91035, Israel
iaakov@jce.ac.il

**Abstract.** *Linear Software Models* is a systematic effort to formulate a theory of software systems neatly based upon standard mathematics, viz. linear algebra. It has appeared in a series of papers dealing with various aspects of the theory. But one was lacking a single source for its key ideas. This paper concisely distills foundational ideas and results obtained within *Linear Software Models*. First and foremost we claim that one must have a deep comprehension of the theory of software – the aim of this effort – before embarking into theory and practice of software engineering. The Modularity Matrix is the central algebraic structure of the software theory: it is the source of quantitative modularity criteria; it displays high cohesion, i.e. high sparsity; a Standard Modularity Matrix is defined – square and block-diagonal – enabling designs comparison for any software systems in a Software Design Laboratory. It triggers formulation of novel open questions waiting for resolution.

## 1  Introduction

The goal of this paper is to be a single source for the key ideas of *Linear Software Models*. It concisely summarizes and reformulates concepts and results obtained within *Linear Software Models*. The main motivation is to clarify essential issues that were raised in discussions following workshops and conference presentations, while distilling its foundational ideas. To this end, it does offer new concepts and alternative proofs of some theorems. It contains fundamental concepts, essential theorems, basic equations and calculation procedures. It does not include case studies, detailed numerical tables of results and extensive bibliography of related work. For the latter material the reader is referred to the other papers in the series. This paper is expected to gradually evolve into new updated and more precise versions.

## 1.1 Assumptions

We assume that the reader is well-versed in object oriented programming languages and design. For instance, we do not define or explain:

- "classes" and other common terms found in such languages;
- software "design patterns";
- extensions to classes, such as "aspects".

No previous knowledge is assumed about *Linear Software Models*.

## 1.2 The Structure of this Paper

This paper contains numbered items (notation given below within parentheses):

- Concepts – Undefined but explained (C1, C2, …)
- Definitions – using previous concepts and algebra (Def1, Def2, …)
- Theorems – using definitions and algebra (Theorem1, Theorem2, …)
- Equations – (Eq1, Eq2, …)
- Frequent Questions – (FQ1, FQ2, …)
- Open Questions – (OQ1, OQ2,…)
- Procedures – (Proc1, Proc2, …)

The organization of the remaining of this paper is as follows. Each section contains just one foundational idea, which is developed using its concepts, theorems, equations and procedures. Section 2 describes the relation between a theory of software and software engineering. Section 3 deals with the Modularity Matrix as the basic algebraic structure expressing modularity. Section 4 presents the idea of a theoretical Standard Modularity Matrix being square and block-diagonal. Section 5 deals with the Modularity Matrix as the source of quantitative criteria for modularity. Section 6 presents the formalization of intuitive software notions – e.g. cohesion – in terms of the Modularity Matrix. Section 7 deals with the Software Design Laboratory, where comparisons are made between different designs of any software system, and the Standard Modularity Matrix. In particular an open question on eventual Bordered Modularity Matrices is formulated.

The above sections are concluded by a discussion of the foundational ideas (section 8). At the end of the paper there is a brief annotated bibliography of the Linear Software Models' papers (section 9), followed by a concise bibliography of related papers of interest to the Linear Software Models (section 10). Finally, Appendix A illustrates addition operations on Modularity Matrix vectors.

# 2  Deep Understanding of Software Precedes Software Engineering

Our first claim is that a deep understanding of software theory should precede the theory and practice of software engineering.

## 2.1  A Theory of Software

*Linear Software Models* is a theory of software systems dealing with software composition. We assume that software systems have hierarchical structure. A system is composed of sub-systems, which in turn are composed of sub-sub-systems, down to the lowest level of indivisible architectural units. A central concept in the theory of software composition is modularity.

*Linear Software Models* is an effort to formulate a rigorous, simple and clear Theory of Software based upon well-established mathematics, viz. linear algebra. This justifies the *Linear* name given to these Models. We believe that it is an achievement to use known and well-founded mathematics, without the need to invent and justify otherwise arbitrary mathematical tools. Thus, our whole effort is to characterize software concepts in terms of mathematical notions, while enabling software intuitive interpretations.

A comprehensive theory of software systems has additional properties besides algebraic composition – such as semantic properties. This paper focuses only on algebraic composition. Semantic properties will be dealt with in other papers.

## 2.2  Software as a Basis for Software Engineering

A theory of Software is a necessary, independent basis for theory and practice of software engineering.

We understand that the current mainstream field of software engineering includes topics beyond the software system, which is itself described by the theory of software. These other software engineering aspects are mainly related to the humans involved with software: stakeholders that order software; individual software engineers which develop and maintain the software systems; teams of developers and their interactions as relevant to production of quality software; methods and processes used by human developers, say "agile" methods; economic and societal implications of software.

This paper does not refer to all these software engineering aspects. These would be dealt with by a theory of software engineering, which is complementary to the theory of software in this paper.

# 3  The Modularity Matrix: an Expression of Modularity

The idea of modularity for a Software System is that it is composed of sub-systems which are independent of each other. We postulate independence to mean "linear independence" within linear algebra. Therefore, the Modularity Matrix is a natural way to express and analyze modularity.

## 3.1  Software Architectural Units

Any software system is conceived and built with a purpose, say to calculate some functions – e.g. for weather forecast, to simulate a system of elevators in a building or to perform remote robotic surgery.

The design of a software system is formulated in terms of architectural units. Architectural units represent system structure and system functionality. Its basic concepts are:

C1 – **Structors** – these are structural units which describe software architecture. Typical structors are – structs, classes, interfaces, aspects – and their collections (for instance sets of classes, as in design patterns).
The basic ideas behind structors are:
- a- *Types not instances* – Structors refer to types not to instances;
- b- *Generalization of structures* – Structors are a generalization of structures to any hierarchical level of a software system; thus they deserve a special name "structors" – and not just say "classes";
- c- *Finite algebraic vectors* – Structors are represented by finite algebraic vectors, and their name should remind us of vectors.

C2 – **Functionals** – these are behavioral units which describe software architecture. Typical functionals are – functions or methods of classes, roles in design patterns (as given by Riehle [25]), families of related functions (e.g. trigonometric or hyperbolic functions).
The basic ideas behind functionals are:
- a- *Potential functions* – Functionals are *potential functions* within a software system that can be, but are not necessarily invoked;
- b- *Generalization of behaviors* – Functionals are a generalization of behaviors to any hierarchical level of a software system; thus they deserve a special name "functionals" – and not just say "functions";
- c- *Finite algebraic vectors* – Functionals are also represented by finite algebraic vectors.

C3 – **Modules** – these are architectural units within a software system, composed of grouped structors and their corresponding grouped functionals. The basic ideas behind modules are:
- a- *Higher than basic level* – Modules are architectural units in a higher abstraction level than the basic structors and functionals of a system hierarchy;
- b- *Recursively defined* – Modules are a generic term, which can be recursively defined; thus we have modules of modules, up to the whole system.

## 3.2   Deployable Components

Deployable units are also composed of structors and functionals. But the design logic of manufacturers of deployable units is different from software system design. The common logic of deployable units is their manufacturer's core technology. The basic concept is:

C4 – **Components** – these are deployment units.  These are often purchased as COTS (Commercial Off-The-Shelf) components from different manufacturers. Typical components nowadays are programming language defined – a Java archive (jar), a C# assembly, a C++ dynamically linked library (dll).

The basic ideas behind components are:

a- *Indivisible deployable units* – Components are purchased as indivisible deployable units; as such, not all structors/functionals of a component may be required by a software system; some functionals may never be invoked;

b- *Diverse sources* – Components from different manufacturer sources may be diverse.

## 3.3   The Software Composition Problem

The software composition problem is how to design a modular software system from available components, either built in-house or purchased as COTS components. Note that the latter are usually not designed specifically for a particular system. Thus:

C5 – **Software Composition Problem** – this problem is formulated as follows: given a list of structors and a list of functionals that are required in the system, choose a minimal number of structors that satisfy the system requirements.

The basic ideas behind the ***solution*** of the software composition problem are:

a- *Choose linearly independent sets* – The solution of the software composition problem, within a Linear Software Model, is to choose linearly independent sets of structors and linearly independent sets of functionals;

b- *Modularity Matrix manipulation* – The algebraic technique used to solve the software composition problem is, first to represent the software system by a matrix – the Modularity Matrix – and then apply known methods of matrix manipulation.

## 3.4   The Modularity Matrix

The Modularity Matrix is a linear algebra structure representing a software system, enabling analysis of the system design, and its modularity. It is defined as follows:

Def1 – **Modularity Matrix** – A fully expanded Modularity Matrix is a matrix with real numbers, restricted to binary values, asserting links (1-valued elements) between software functionals (row vectors) and software structors (column vectors). The absence of a link is marked by a 0-valued element.

The basic ideas behind the Modularity Matrix are:

a- *Basic structors are indivisible* – Basic software structors are assumed indivisible into smaller structors, although decomposition could be possible in principle; similarly basic software functionals are assumed indivisible into smaller functionals, unless they are explicitly named as compositions (say the family of trigonometric functions);

b- *Structor provides Functional* – A 1-valued element linking a functional to a structor means that the structor provides that functional; it may be that more than one structor provides that same functional; for instance, it is conceivable that one structor may provide the direct trigonometric functions (sin, cos, …) and another structor may provide the inverse trigonometric functions (arcsin, arccos, …), such that together these two structors provide the "family of trigonometric functions"; any kind of "global" notion (that may exist in certain programming languages) would be a property of the whole system and not of specific modules, thus not touching composition and not appearing in the Modularity Matrix;

c- *Inheritance provides same Functional* – Another case in which two structors provide the same functional is inheritance between a class and its super-class (or alternatively an interface in certain programming languages); in this case the super-class may provide the declaration of a functional and the class may provide the actual functional that can be invoked;

d- *Linearly Independent structors/functionals* – Different structors must be represented by distinct vectors; different functionals must also be represented by distinct vectors; the generic criterion for independent structors/functionals in any subsets of a system is linear independence;

e- *Real numbers with real operators* – In spite of the restriction to binary values in the above definition Def1, the Modularity Matrix in its most generality has matrix elements as real numbers with real operations – as will be justified later on in this paper. In principle, one could have even matrix elements with complex values. The latter case is still not discussed in this paper.

f- *Arithmetic Operations on Modularity Matrix elements* – there are several reasons for doing arithmetic operations on the matrix elements of the Modularity Matrix or on matrices derived from the Modularity Matrix.

To begin with, one wishes to obtain linear combinations such as:

$$LinComb = c_1 * x + c_2 * y + c_3 * z \qquad (1)$$

where $c_i$ are constant coefficients and $x, y, z$ are elements of a set, say matrix elements or vectors. Thus the relevant arithmetic operations are:

- *addition* of matrix elements;
- *multiplication by a scalar* number.

Typical calculations that can be built from these arithmetic operations are:

- test linear independence;
- calculate inner products;
- obtain eigenvectors/eigenvalues;
- multiply matrix elements by suitable weights;
- calculate various functions, such as diagonality.

There are two questions that one may ask:

1- What is the interpretation of these operations? We do have interpretations for certain operations; e.g. summing two matrix elements found in two structors and in the same functional gives the number of times the functional is provided by a module containing both structors (see Appendix A at the end of this paper). But we adopt the position that intermediate calculations may not always have an interpretation, as long as the final calculation is understood.

2- Are there limitations on these operations? The answer is that there may be limitations: we prefer not to multiply Modularity Matrix elements by negative numbers. But here again, following the custom of mathematical theories of experimental sciences, one may perform formal intermediate calculations, as long as the final calculation is meaningful.

# 4  Standard Modularity Matrices: Square and Block-Diagonal

*Linear Software Models* is a systematic effort to formulate a theory of software systems. Therefore it proposes a theoretical standard against which designs for each specific software system should be compared. Our perspective is that of an experimental science in which the proposed and accepted theory is compared with experimental measurements in a Software Design Laboratory, for each design of a specific software system.

Assuming that the theory is correct, and the design for a system deviates from the standard, one should improve the design until it approaches the standard. The optimal system design should be the best approximation to the standard.

In the eventual situation that reasonable optimized designs for a large variety of systems, along extended experimentation periods, disagree with the theoretical standard, the theoretical model should be improved.

The theoretical standard proposed by *Linear Software Models* for each software system is a square and block-diagonal Modularity Matrix for the respective system.

## 4.1  Linear Independence of Structors and Functionals

Preliminary definitions, needed to formulate theorems on the characteristic of standard Modularity Matrices, refer to independent structors and functionals.

Def2 – **Independent Structor** – A software structor is independent of other structors in the system, if it provides a non-empty proper subset of functionals of the system, given by the 1-valued links in the respective column, and is linearly independent of other columns in the Modularity Matrix.

The basic idea behind independent structors:

a- *Non-empty proper subset* – The proper subset requirement stems from the fact that a column totally filled with 1-valued elements would not be a characteristic of a particular structor, but from the whole system.

Def3 – **Independent Functional** – A software functional is independent of other functionals in the system, if it corresponds to a non-empty proper sub-set of system structors, given by the 1-valued links in the respective row, and is linearly independent of other rows in the Modularity-Matrix. This set is the composition set of the functional. The basic ideas behind independent functionals are mutatis mutandis the same as those behind independent structors.

## 4.2  The Square Well-Composed Modularity Matrix

The standard Modularity Matrix for each software system is square as shown by the next theorem.

Theorem1 – **Square Modularity Matrix** – If in a Modularity Matrix all its structors are linearly independent and all its functionals are linearly independent, the Modularity Matrix number of structors $N_S$ is equal to its number of functionals $N_F$. The matrix is Square.
Such a matrix is called a *Well-Composed* Modularity Matrix. An alternative proof is given here, differing from that in previous references (e.g. in [10]):

*Proof:* Assume the Modularity Matrix has no empty (all zero-valued) rows/columns, and the Modularity Matrix has no completely full (all 1-valued) rows/columns.

The column space of the Modularity Matrix is the subspace $\mathbf{R}^m$ spanned by its structors (the columns). The row space of the Modularity Matrix is the subspace $\mathbf{R}^n$ spanned by its functionals (the rows). $\mathbf{R}^m$ and $\mathbf{R}^n$ are written with distinct superscripts, as these subspaces are not assumed to have the same dimension.

Since by the theorem hypotheses all matrix structors are linearly independent and they span the column space, the structors are a basis for the column space. Mutatis mutandis, all matrix functionals are linearly independent and they span the row space, thus the functionals are a basis for the row space. By the "Fundamental Theorem of Linear Algebra" (see references Strang [27] and Stover [26]) the column space and the row space of a matrix both have the same dimension. In other words: the number of independent columns equals the number of independent rows and equals the matrix rank.                                                                                □

The last proof sentence is enough as a whole proof. But we need column/row spaces for later discussion of Modules coupling (cf. sub-sections 6.1 and 6.4).

The basic idea of this Square Modularity Matrix theorem is:

a- *All structors/functionals are respectively linearly independent* – The actual number of functionals in a software system may be different from the number of structors; only if all functionals are linearly independent and all structors also are linearly independent, according to the theorem hypotheses, the matrix is square (see FQ1 below for an extended discussion of this recurring issue).

### 4.3   The Block-Diagonal Modularity Matrix

The standard Modularity Matrix for each software system is also block diagonal as shown by the next theorem. We denote the union of *composition sets* for a set of functionals as "*union set*".

**Theorem2 – __Block-Diagonal Modularity Matrix__** – Any well-composed Modularity Matrix in which the union-set for a given set of functionals is disjoint to the other functional union-sets is *reducible*, i.e. it can be put in block-diagonal form.
A proof outline is given in ref. [10].

The basic idea of this Block-Diagonal Modularity Matrix theorem is:

a-  *__Diagonal blocks are disjoint__* – The essence of modularity is lack of dependencies among modules allowing system composition by additivity of modules.

### 4.4   The Standard Modularity Matrix

By the previous theorems the standard Modularity Matrix for each software system is:
- *Square* – same number of linearly independent structors and linearly independent functionals;
- *Block diagonal* – with square blocks (smaller than the whole matrix) along the diagonal, and zero-valued matrix elements outside the blocks.

A Standard Modularity Matrix for a software system dealing with calculations and display of geometric shapes is seen in Fig. 1.

| Structors → | | Circle | Triangle | Shape | GUI | Refresh Aspect |
|---|---|---|---|---|---|---|
| **Functionals ↓** | | S1 | S2 | S3 | S4 | S5 |
| **Calculate-Circle-Functions** | F1 | 1 | 0 | 0 | 0 | 0 |
| **Calculate-Triangle-Functions** | F2 | 0 | 1 | 0 | 0 | 0 |
| **Translate-Shape** | F3 | 1 | 1 | 1 | 0 | 0 |
| **Display** | F4 | 0 | 0 | 0 | 1 | 0 |
| **Refresh** | F5 | 0 | 0 | 0 | 1 | 1 |

**Fig. 1.** *Geometric Shapes Standard Modularity Matrix*. Structors are column vectors and functionals are row vectors of the Matrix. This Modularity Matrix is square (5*5) and strictly block-diagonal. The two blocks (with blue hatched background) are the modules of this system. The top-left module refers to geometric shapes. The right-bottom module refers to the Graphical User Interface (GUI) and its operations (Display and Refresh). Refresh operations, activated after every new shape display or translation of an existing shape, are collected in a Refresh Aspect. All matrix elements outside the modules are zero-valued. The Translate-Shape functional is declared by the Shape Structor and inherited by both specific geometric shapes.

## 4.5   Why is the Standard Modularity Matrix Square?

A recurring issue when discussing Linear Software Models is the question: why the Standard Modularity Matrix is square? Here we give intuitive interpretations as replies to this issue.

**FQ1 – <u>Square Modularity Matrix</u> –** A frequent challenging question: is the number of functionals restricted at all by the number of structors? One could easily imagine and actually add more functions to classes in reasonable ways, breaking the hypotheses of Theorem 1, i.e. breaking the condition of a square Modularity Matrix.

The exact formal reply given by theory is just one: the square consequence of Theorem1 is ***only true*** for "all structors of a matrix being linear independent" ***and*** "all functionals of the same matrix being linear independent". If these conditions are not imposed and obeyed for a given system design, one may obviously have as many functionals as one wish for any number of structors.

There is no other reply given by the theory. To facilitate understanding we may give interpretations for the Modularity Matrix being square under the above conditions. These interpretations are not to be taken as a substitute for the formal reply given by the theory. Here they are:

a-  <u>Functionals are coalesced if</u> they have identical rows in the matrix. When does this happen? This is explained in the next items.

b-  <u>Functions having closely related meanings</u> – for instance, the functions *attach* and *detach* which occur in the "Subject" class of the Observer design pattern mean that the subject may add (attach) a new "Observer" to its list of observers, or delete (detach) an existing "Observer" from its list. Thus *attach* and *detach* are inverse operations and have identical rows in the matrix. One could even write:        $detach = attach^{-1}$
Thus we either may coalesce both functions in the same row or substitute both operations by a single one meaning "maintain list of observers".

c-  <u>Functions depending on a common set of variables</u> – for instance, the "Calculate-circle-functions" in Fig. 1 could be: calculate the circle area, calculate the circle perimeter, calculate the circle diameter, etc. All of these depend on ***r*** "radius" of the circle. The circle area $= \pi r^2$; the circle perimeter $= 2\pi r$; the circle diameter $= 2r$; etc. This explains why all these functions have the same composition dependence on the class circle.

d-  <u>The ideal modularity is strictly diagonal</u> – The ideal block-diagonal matrix is the strictly diagonal. This is the simplest case of modularity as it has a bi-directional one-to-one relation between structors and functionals. In this kind of matrix one obviously has equal numbers of structors and functionals.

e-  <u>Kinds of dependencies without repetition</u> – the Standard Modularity Matrix only shows explicitly all the different kinds of dependencies (of functionals on structors) *without repetition*.

# 5   The Modularity Matrix: the source of Quantitative Modularity Criteria

An important consideration for software system design analysis is to use the Modularity Matrix itself as a source of quantitative criteria to evaluate the modularity of a given design for a software system.

This in contrast to using an additional external model superimposed on the Modularity Matrix to obtain the referred criteria. Using the Modularity Matrix itself for this purpose has two advantages:

a- simplicity – to base all calculations on a single basic algebraic structure, instead of using two or more models of different nature;

b- self-consistency – no need to provide justifications for the arbitrary combination of models.

An example of combining different models is the use – in addition to the Modularity Matrix of the software system – of an economic model of software development, taking into account activities of developer teams and the resources involved. Such combinations do not belong to the software system theory description. They rather fit the complementary theory of software engineering (mentioned in sub-section 2.2).

## 5.1   Diagonality

The diagonality of a Modularity Matrix $M$ is taken as a quantitative expression of the software system modularity.

Eq1 – **Diagonality** – The diagonality is the difference between the usual *Trace* – the sum of the Modularity Matrix diagonal elements – and *Offdiag*, a new term expressing off-diagonal elements:

$$Diagonality(M) = Trace(M) - Offdiag(M) \quad (2)$$

The basic ideas behind Diagonality are:

a- *Simplest case is strictly diagonal* – since the standard Modularity Matrix is block-diagonal and its simplest case is strictly diagonal, we measure the modularity by a distance criterion from the strictly diagonal Matrix (see below the definition of Offdiag);

b- *The more positive the diagonality* – the 1st term of the Diagonality is always positive; the 2nd term is always negative as the offdiag is always positive, but has a negative sign; the more positive the Diagonality, the more modular a particular design of the software system under analysis.

c- *Integer number calculation* – This equation is a first example of a non-binary number calculation based upon the Modularity Matrix; the inputs are *"binary"* values 1.0 and 0.0, but the output is an integer number (either positive, zero or negative); actual real numbers will be used in later equations.

## 5.2   Offdiag

Eq2 – **Offdiag** – Offdiag is a sum over 1-valued $M_{jk}$ off-diagonal elements of the square $N*N$ Modularity Matrix $M$, multiplied by the absolute value of the difference between each element's row $j$ and column $k$ indices. The overall matrix offdiag is a double sum over rows and columns:

$$offdiag(M) = \sum_{j=1}^{N} \sum_{k=1}^{N} M_{jk} \cdot \mid j - k \mid \qquad (3)$$

The basic ideas behind *Offdiag* are:
a- <u>*Calculation simplicity*</u> – the choice of $\mid j - k \mid$, called Manhattan, city-block or taxicab distance was dictated by calculation simplicity;
b- <u>*Distance between matrix elements and diagonal*</u> – $\mid j - k \mid$ directly reflects distances between off-diagonal matrix elements and the Modularity Matrix diagonal, in which the distance is zero;
c- <u>*Offdiag depends on reordering*</u> – Obviously, *offdiag* is dependent on the reordering of rows and columns; minimizing offdiag implies approaching the standard Modularity Matrix, which is block-diagonal.

# 6   Formalizing the Common Wisdom of Software

One of the success' tests of a theory is whether it provides a formal basis to widely accepted intuitive notions. *Linear Software Models* indeed offer a formal basis to the common wisdom of software.

Here we translate a series of informal notions into naturally formalized concepts. These are grouped into two sets with respect to the Modularity Matrix:

- ***2-dimensional concepts*** – referring to matrix modules, either their internal or external characteristics, such as *modules* proper, *cohesion*, *composition*, 2-D-*coupling* and *single responsibility*.
- ***1-dimensional concepts*** – referring to matrix rows or columns, such as 1-D-*coupling* and *inheritance*;

## 6.1   Modules

Def6 – **Modules** – modules (previously characterized as an undefined concept in C3) are now formally defined as the disjoint diagonal blocks of structors/functionals in the Modularity Matrix of a software system.

The basic ideas behind *modules* are:
a- <u>*Modular system*</u> – the Modularity Matrix corresponds to the intuitive idea of modular software systems;
b- <u>*At least two disjoint blocks*</u> – a Modularity Matrix with a single block is not a module, but just the whole system. The condition for a system to have modules is at least two disjoint blocks in the Modularity Matrix.

This condition closely corresponds to the requirement of a proper sub-set of functionals for independent structors;

c- *Block operations* – *Collapse* and *expansion* operations on the Modularity Matrix unfold a whole hierarchy of block levels, where a level is characterized by the expanded blocks at that level. *Block collapse* reduces a whole block into a single matrix element (a black-box with invisible internal matrix elements). *Block expansion* restores the collapsed black-box to a white-box with visible internal matrix elements;

d- *Number of modules at a level* – is the number of diagonal blocks in the partition of the Modularity Matrix at the given level;

e- *Sub-spaces* – since all structors within a module are linear independent, and they span the column space of the module block matrix, the module structors are a basis for this module column space; mutatis mutandis for module functionals and its row space; the same is true for the whole Modularity Matrix structors: thus, the (structor) column space of the whole Modularity Matrix is additively composed of the (structor) column spaces of the individual modules; mutatis mutandis for whole Modularity Matrix functionals and its row spaces.

## 6.2   Cohesion

Def7 – **Cohesion** – cohesion within a module formally means that the links among architectural units – structors and functionals – are restricted to the units of the given module and do not extend to other sub-sets of units.

The basic ideas behind *cohesion* are:

a- *Sparsity* – an important property of the Modularity Matrix of software systems, which is a consequence of cohesion, is that the matrix is sparse – i.e. large numbers of zero-valued matrix elements relative to the matrix size. Sparsity of a module or of the whole Modularity Matrix can be quantitative criteria within algorithms and to verify Modularity Matrix trends with size;

b- *Modularity Matrix maximal sparsity* – the Modularity Matrix of given size M with maximal sparsity is the strictly diagonal matrix – i.e. the unit matrix of size M {see equation (4) in Appendix A, in page 208 of ref. [10]};

c- *Sparsity increases with size M* – the sparsity of a Modularity Matrix with size M, increases with M (see Appendix A, in page 208 of ref. [10]).

## 6.3   Composition

Def8 – **Composition** – the specific meaning of composition of structors (classes) within a module is the formation of a module, while avoiding the coupling of structors (including inheritance) within the given module.

The basic ideas behind *composition* are:

a- *Specific to structors* – refers to structors within a module as opposed to the generic *Software Composition Problem* characterized in the undefined concept C5;

b- *Avoid Coupling* – means that less 1-valued matrix elements are found in the same row or same column within a module;

## 6.4   2-D Coupling

**Def9 – <u>2-D Coupling</u>** – Two-dimensional coupling refers to coupling between modules. Coupling between two or more modules means that structors or functionals of a module of a given Modularity Matrix are not orthogonal respectively to structors or functionals of another module of the same matrix.

The basic ideas behind *2-D Coupling* are:

   a- *Coupling by outliers* – 2-D coupling among modules is caused by one or more outliers – i.e. 1-valued matrix elements outside the diagonal blocks;

   b- *Sub-spaces* – if there are one or more outliers: the structors within each of the modules still span the column space of the respective module (they exclude the outliers for the respective modules); mutatis mutandis for the functionals and its row space; but with 2D-Coupling, one cannot exclude the outliers from the whole Modularity Matrix (structor) column space: if structors are not orthogonal, but still are linear independent (this is possible!) the structors still span the whole matrix column space; on the other hand, the structor column space of the whole Modularity Matrix is *NOT* anymore additively composed of the module structor column spaces; mutatis mutandis for the functionals and the whole matrix row space.

## 6.5   Single Responsibility

The so-called Principle of Single Responsibility (see e.g. Martin [21]) is now a theorem upon a block-diagonal Modularity Matrix. Given its less importance we could have called it just a Lemma

**Theorem3 – <u>Single Responsibility</u>** – In a standard (square and block-diagonal) Modularity Matrix each structor (column) intersects a single module. Similarly each functional (row) intersects a single module.

The theorem proof is straightforward.

The basic ideas behind *Single Responsibility* are:

   a- *Functionally Related Structors* – a single module is responsible for providing each functional exclusively by its related structors. In other words, functionally related structors are grouped in a single module.

## 6.6   1-D Coupling

**Def4 – <u>1-D Coupling</u>** – Coupling among 1-dimensional architectural units – structors or functionals – means that there are linear dependencies among sets of respectively columns/rows of the Modularity Matrix of the software system.

The basic ideas behind *coupling* are:

   a- *The informal idea of mutually dependent components* – see e.g. what the GoF book [14] states in its Glossary (page 360): "coupling is the degree to which software components depend on each other". This is stated without specifying the meaning of "degree of dependence".

b- *Operational decoupling* – the operational meaning is the removal of linear dependencies, by some suitable action. This action may be addition or removal of architectural units and/or links between units.

## 6.7   Inheritance

Def5 – **Inheritance** – Inheritance of a structor (an interface or a class by another class) means that a single functional (row) is common to two or more structors (columns). This implies a set of two or more 1-valued Modularity Matrix elements in a single row of the matrix.

The basic ideas behind *inheritance* are:

a- *A type of coupling* – inheritance is a type of coupling of two or more structors by a functional. This is the reason for the GoF book [14] advice in page 20: "Favor object composition over class inheritance."

b- *A legitimate type of coupling* – despite coupling being a negative characteristic, inheritance is considered a legitimate type of coupling; legitimacy has been imparted by its systematic use in design patterns [14].

c- *Inverse is not True* – a set of two or more 1-valued Modularity Matrix elements in a single row of the matrix, can be, but is not necessarily a case of inheritance. Counter-example: a composite functional (e.g. the family of trigonometric functions is composed of the direct and the inverse functions), as we have seen in the ideas behind the Modularity Matrix (in page 6 of this paper).

d- *Liskov Substitution Principle* – inheritance has been defined by the Liskov Substitution Principle [19]. This kind of definition is interesting, but not so formal: it certainly deserves discussion. It is further formalized in [20].

e- *Explicit syntax* – inheritance of a class from an interface/class is explicit in the syntax of certain object oriented programming languages. But, explicit syntax is not a substitute for a formal definition.

As an example of inheritance, in Fig. 1 the "Translate-shape" functional (F3) is inherited by sub-classes 'circle' (S1) and 'triangle' (S2) from the parent class 'shape' (S3).

# 7  Experimental Theory Corroboration in the Software Design Laboratory

As stated earlier in this paper – in the beginning of the presentation of the Standard Modularity Matrices theory in section 4 – our perspective of software is that of an experimental science in which the proposed and accepted theory is compared with experimental measurements in a **Software Design Laboratory**, for each experimental design of a specific software system.

In this section, we characterize a Software Design Laboratory, and then we shortly review representative results of experimental theory corroboration.

## 7.1 The Software Design Laboratory

The *Software Design Laboratory* is characterized by the explicit use of modularity as essential to software system design. Typical modularity design activities are:

    a-  *Preparation of system software Modularity Matrix* – starts either by manual preparation of structors and functionals lists from system requirements and available COTS and/or by means of a software tool, given existing sub-systems;

    b-  *Designs comparison by quantitative modularity criteria* – Comparison of two or more proposed designs for a given software system by means of quantitative modularity criteria;

    c-  *Measurement against Standard Modularity Matrix* – Checking a software system design against the corresponding Standard Modularity Matrix obtains eventual outliers and their respective quantitative expressions of lack of modularity; this is expected to serve as a basis for future modularity tools that will (semi-)automatically reduce couplings and simplify software design;

    d-  *Coupling Diagnosis based upon outliers* – given outliers, locate eventual couplings to be solved; solution by addition/deletion of structors/functionals; this may affect more than one existing module;

    e-  *System redesign* – overall software system or sub-system redesign, stimulated by the Modularity Matrix appearance and/or by outlier coupling solutions.

It should be clear that explicit use of modularity cannot be based solely on programming/debugging at the programming language level. Individual programming language commands do not directly reflect the system design modularity characteristics. This is not to be misunderstood as meaning that one may neglect programming/debugging techniques or the latter techniques should be done in separate from software modularity design. Modularity design comes above the usual programming language level.

Explicit use of modularity in software design needs dedicated tools to make software modularity computations and software design manipulation such as the above modularity design activities.

## 7.2   Strictly Linear Software Models

We started the corroboration efforts – or rather falsification efforts (in Popper's falsifiability [29] terms) – in the first papers in this series. Unless the theory or its eventual concept expansions (see e.g. Buzaglo [4]) breakdown, these efforts should be continuous.

Among the positive corroboration results, there were two kinds of software sub-systems collected under the name of **Canonical Sub-systems** that were shown to strictly obey *Linear Software Models*. The first kind consists of sub-systems explicitly built to explain the idea of modularity, thus being canonical in this sense. Parnas'

KWIC index [23] is such a case. The second kind consists of collections of small sub-systems claimed to be frequently used, and recommended to be part of the software engineering basic vocabulary, thus also being canonical in this latter sense. Software *Design Patterns* (e.g. in the GoF book [14]) are included in this kind.

In a posteriori critical review of these positive corroboration results, the Modularity Matrices that we constructed of Parnas' KWIC index modularizations seem to be neat and strictly correct. Referring to Design Patterns as canonical sub-systems, it is also claimed that given the choices of a specific class diagram and its participants for a given design pattern (extracted from the GoF book) and the respective list of structors and functionals, the Modularity Matrices seem to be neat and strictly correct. The results for *Design Patterns* could generate some controversy, due to the lack of formality of UML and the multitude of inconsistent published class diagrams for design patterns, including in the GoF book itself, as was noted in the discussion of our paper [12]. But we must stress that such inconsistencies are not due to inherent problems of Modularity Matrices. We summarize the positive experimental results, by saying that they indeed corroborate the theory foundations.

### 7.3   Bordered Linear Software Models

Regarding larger real software systems, samples found in the internet were used to build and test their Modularity Matrices. An example of a Modularity Matrix is seen in Fig. 2 for the NEESgrid system for "Network Earthquake Engineering Simulation" [13]. Detailed considerations can be found in papers of this series (e.g. [10] and [12]).

The systematic outcome for larger systems is called *Bordered* Linear Software Models, i.e there are a few outliers near the diagonal block borders (see Fig. 2).

| Structor | | Data Str | Data Rp | Data Vu | Data Ac | Tele pre | Chef | Grif Infr | Sim Rep | Hyb Exp | Data Dis |
|---|---|---|---|---|---|---|---|---|---|---|---|
| **Functional** | | S1 | S2 | S3 | S4 | S5 | S6 | S7 | S8 | S9 | S10 |
| CollData | F1 | 1 | 1 | 0 | 1 | | | | | | |
| MngData | F2 | 0 | 1 | 1 | | | | | | | |
| DataView | F3 | 0 | 0 | 1 | | | | | | | |
| OtherCol | F4 | | | 1 | 1 | 0 | 1 | 1 | | | |
| SynCol | F5 | | | | 0 | 1 | 1 | | | | |
| AsynCol | F6 | | | | 0 | 0 | 1 | | | | |
| HPC | F7 | | | | | | | 1 | | | |
| SimCodes | F8 | | | | | | | | 1 | | |
| HybExp | F9 | | | | | | | | | 1 | |
| SercData | F10 | | | | | | | | | | 1 |

**Fig. 2.** *Example of a Bordered Modularity Matrix*. This matrix refers to the NEESGrid system. One can see two 3*3 diagonal blocks and four 1*1 strictly diagonal elements (light blue background). Near the borders of the bigger blocks there 3 outliers (hatched dark blue background). The overall matrix is indeed sparse: all the matrix elements except the diagonal blocks and outliers are zero-valued.

## 7.4   An Open Question: Bordered or Not?

It is an open question, deserving further investigation, whether *Bordered* Linear Software Models is a separate category of systems:

**OQ1 – <u>Bordered Linear Software Models</u> –** Is Bordered Linear Software Models a separate category of software systems? The reason for this question is that: on the one hand, the large systems taken as examples from the internet were developed before the theory of Linear Software Models; on the other hand, now that we developed the theory we do not have enough knowledge about these systems, in order to decouple the outliers and achieve a strictly block diagonal Modularity Matrix.

Possible answers to this question are:

a-   *Bordered Linear Software Models are rare* – these would be negligible exceptional cases, that perhaps deserve some investigation, but do not affect the general case of strictly block diagonal matrices;

b-   *No Bordered Linear Software Models* – once these cases are analysed with full information, one could decouple all the bordered outliers, always resulting in strictly *Linear Software Models;*

c-   *Bordered Linear Software Models are a special category* – if none of the previous cases is true, then the theory must be expanded to accommodate and explain this category; in this third case, one should then ask a further question, whether there are more than two categories (strict and bordered) of Linear Software Models.

# 8  Discussion: Foundational Ideas

This discussion focuses on whether the presented results and ideas are indeed the foundational ideas of the *Linear Software Models*. It is clear that not all ideas have the same weight, in particular referring to the concepts and definitions which are followed by a list of ideas, instead of just one or two key ideas. A partial justification for these lists is the secondary purpose of this paper, viz. as a clarification of difficult points raised in discussions after conference/workshop lectures.

So here we summarize the current shortest set of the foundational ideas:

1.   Deep Understanding of Software before Software Engineering;
2.   The Modularity Matrix;
3.   The Standard Modularity Matrix;
4.   The Modularity Matrix: source of quantitative modularity criteria;
5.   Cohesion;
6.   The Software Design Laboratory;
7.   The Open Question: Bordered or Not?

## 8.1   Foundational Ideas

Here we provide arguments for why these are foundational ideas.

### 1.   Deep Understanding of Software before Software Engineering (SE)

This means three things: a- there is a deep theory of software, of which the linear algebra formulation is only one aspect of it; software semantics is another aspect of it, not dealt with in this paper; b- it is important to clearly separate the theory of software from other aspects of Software Engineering – say human, social and economic – in order to achieve deep understanding of software (see a similar argumentation in e.g. Perry & Batory [24]); the Modularity Matrix is entirely contained and justified within a purely software theory, in contrast to other kinds of matrices which need e.g. economic considerations to obtain modularity criteria; c- deep understanding of Software before jumping into the troubled methodological waters of SE should greatly simplify SE.

### 2.   The Modularity Matrix

The rationale is: a- The Modularity Matrix is the basic algebraic structure within the Linear Software Models. As far as possible any other results and procedures are obtained by directly using it; b- The Modularity Matrix combines structure architectural units (the columns) with behavior architectural units (the rows); c- The whole hierarchy of the software system – at the various abstraction levels – is representable in the Modularity Matrix: the system, the sub-systems, the sub-sub-systems, and so on.

### 3.   The Standard Modularity Matrix

The rationale behind this standard is: a- the Standard is a theoretical result, enabling comparison of the theory with experiments; b- since the columns and rows represent different quantities (structure vs. behavior), being square is not a trivial property of the Modularity Matrix (in contrast to matrices which have the same quantities in columns and rows); c- The Standard Modularity Matrix (square and block-diagonal) is valid for any conceivable software system; thus each design for a given software system can be compared with the Standard; d- the deviations of a design from the Standard point out where are the problems to be decoupled; e- the Standard Modularity Matrix has a simple form that can be easily captured by visual inspection (besides performance of calculations).

### 4.   The Modularity Matrix: the source of quantitative Modularity criteria

The important idea is that a single algebraic *Linear Software Model* based upon a central algebraic structure – The Modularity Matrix – is itself the source of quantitative modularity criteria. No need to artificially impose and no need to justify an external model upon the Modularity Matrix to evaluate it. The criteria values are directly computed and justified from the matrix properties themselves.

## 5.    Cohesion

Among the formalized software concepts, *cohesion* is somewhat surprising as a foundational concept. It has two important properties that follow from the fact that the Modularity Matrix sparsity is a consequence of cohesion: a- one can deduce quantitative criteria from sparsity; b- it is easily perceived by visual inspection (besides performance of calculations).

## 6.    The Software Design Laboratory

The Software Design Laboratory embodies the typical duality of experimental science: the interplay between Software theory and laboratory experiments with specific software systems' design. Since the Software theory (of this paper) refers to modularity, the Software Design Laboratory should make explicit use of modularity in software design. It is clear that explicit use of modularity must apply programming/debugging techniques at the programming language level, but it cannot be solely based upon the latter techniques, as individual programming language commands do not directly reflect the system design modularity characteristics. Dedicated Modularity tools are needed for the Software Design Laboratory.

## 7.    The Open Question: Bordered or Not?

One of the benefits of having a software theory is to be able to ask novel definite questions that have no trivial answer. The question whether Bordered Linear Software Models is a distinct category of software systems by itself is an important open question. The answer will demand a combination of laboratory experiments upon a variety of systems, with an eventual theoretical hypothesis to be proved by some formal technique.

# 9  Annotated Bibliography of Linear Software Models

This is a comprehensive annotated bibliography of all papers/videos that appeared in the Linear Software Models series up to the publication of the current paper (from now on referred as the "Key Ideas" Paper = "KeyP").

This version of the KeyP paper is based upon the ideas of the papers already published in journals – and the respective preceding papers presented in conferences and workshops. Papers in the list, which were accepted for journal publication and/or were presented only in conferences and workshops, but not yet published in journals, are not the source of material for the current version of the KeyP paper.

The **A**nnotated **B**ibliography is:

**AB1.** "Linear Software Models for Well-Composed Systems", ICSOFT'2012, ref. [5]. {The first paper in the series, presents definitions, the Modularity Matrix, the Standard Modularity Matrix, and some case studies}.

**AB2.** "Linear Software Models", in GTSE 2012, ref. [6]. {The paper in this GTSE 2012 Workshop proceedings is an extended abstract with contents overlapping the previous ICSOFT'2012 paper; see also next video}.

**AB3.** "Linear Software Models", video of preceding paper, ref. [7]. {This video is a convenient short introduction to Linear Software Models; it also contains the beginnings of the discussion on the paper (with two questions, one of them being FQ1 of this KeyP paper)}.

**AB4.** "Linear Software Models are Theoretical Standards of Modularity", in Vol. 411, CCIS, Springer, ref. [8]. {An extended version of the ICSOFT'2012 paper [5], which was selected to be published in this post-conference wolume}.

**AB5.** "Linear Software Models – Vector Spaces for Design Pattern Modules", in ICSOFT'2013, ref. [9]. {As its title describes, it refers to Vector Spaces for GoF [16] Design Patterns; material not covered yet in this KeyP paper}.

**AB6.** "Linear Software Models: Standard Modularity Highlights Residual Coupling", IJSEKE (2014), ref. [10]. {1[st] journal paper on Linear Software Models covered by this KeyP paper. Contains: basic concepts, definitions, Modularity Matrix theorems, diagonality equations, formalized software concepts, canonical and Bordered Linear Software Model case studies, scalability and applicability of Linear Software Models to real systems, comparison with UML, other matrix models and sparsity calculations}.

**AB7.** "Linear Software Models: Equivalence of Modularity Matrix to its Modularity Lattice", in ICSOFT'2015, ref. [11]. {This paper refers to Modularity Lattices in the Formal Concept Analysis (FCA, see e.g. [15], [16]) approach; material not covered yet in this KeyP paper}.

**AB8.** "Linear Software Models: Decoupled Modules from Modularity Matrix Eigenvectors", IJSEKE 2015, ref. [12]. {Second journal paper on Linear Software Models; material not covered yet by this KeyP paper}.

# 10    Concise Bibliography of Related Papers to Linear Software Models

For a more extensive bibliography of related works the reader is referred to the references in the papers listed in the Annotated Bibliography in the previous section. In this section we just mention a concise (non-comprehensive) sample of books and papers in selected topics relevant to Linear Software Models. These are:

*General Mathematics* – There are innumerous books on Linear Algebra at various levels; a good starting point is Strang's introductory book on Linear Algebra [27]. Specific mathematical results employed in some of the papers of the Linear Software Models include: the fundamental theorem of Linear Algebra (see e.g. Stover [26], the Perron-Frobenius theorem (used in paper [12]), for which one can consult the book by Gantmacher [17] and the notes by Boyle [3].

*General Object Oriented Software* – These include the OMG Unified Modelling Language (UML) specification [22], a well-known overview book on UML [2] and the classical GoF book on Design Patterns [14].

*Alternative matrix-based software design approaches* – Referring to the Design Structure Matrix (DSM) as a different matrix-based software design we mention the "Design Rules" book by Baldwin and Clark [1] whose first volume deals with modularity and a sample paper by Sullivan et al. [28] which also refers to modularity.

*Formal Concept Analysis (FCA)* – Here one finds the relevant background to our paper linking the Modularity Matrix to a Modularity Lattice (Exman and Speicher [12]). This, first of all, includes the books by Ganter and Wille [15] and by Ganter, Stumme, and Wille [16]. The reader is also referred to a sample paper by Lindig and Snelting [18] which assesses modular structure by means of FCA-based techniques.

# Appendix A. Addition operations on Structors and on Functionals

Arithmetic operations on Modularity Matrix elements were mentioned in sub-section 3.4, page 6, among the basic ideas behind the Modularity Matrix. Here we illustrate addition operations by means of simple concrete examples.

Let us look at the addition of two structors belonging to the same module. Assume that both structors provide the same functional, due to inheritance say of a class from another one. The result of the addition is the number of times the functional is provided by the referred module.

In Fig. A1, one sees a fragment of the Modularity Matrix of the Observer design pattern. It shows the addition of "*Concrete observer*" and the "*observer*" structors.

| Structor → | | Concrete observer | Observer | Module= Observer role |
|---|---|---|---|---|
| Functional ↓ | | S4 | S5 | S4 + S5 |
| Maintain list | F1 | 0 | 0 | 0 |
| Notify observers | F2 | 0 | 0 | 0 |
| Maintain global-state | F3 | 0 | 0 | 0 |
| Maintain local-state | F4 | 1 | 0 | 1 |
| Update observers | F5 | 1 | 1 | 2 |
| Display analog | F6 | 0 | 0 | 0 |
| Display digital | F7 | 0 | 0 | 0 |
| Construct objects | F8 | 0 | 0 | 0 |

**Fig. A1.** *Addition of two structors*. Addition of the *Concrete Observer* and *Observer* structors giving the column vector containing the overall provisions in the *Observer role* in the Observer design Pattern. For the functional "Maintain local-state" there is only one provision. For the functional "Update observers" there are *two* provisions due to inheritance: one abstract (provided by the *observer* structor) and one concrete (provided by the *concrete observer* structor). The sum is seen in the right-most column.

Both provide the "*Update observers*" functional, such that the "Observer role" module provides 2 times the referred functional. On the other hand, only the "Concrete observer" structor provides the "Maintain local-state" functional, such that the corresponding addition for the "Observer role" module still gives just 1 provision of the referred functional.

Similar considerations are relevant to addition of functionals (row vectors). The easy interpretation of two 1-valued functionals in a column is that the two of them are provided by a single structor. The addition outcome is the number of individual functionals in the combined functional of the higher-level module, as seen in Fig. A2.

| Structor → | | subject | Concrete subject | Subject resource | Concrete observer | Observer | GUI analog | GUI digit | Init |
|---|---|---|---|---|---|---|---|---|---|
| **Functional ↓** | | **S1** | **S2** | **S3** | **S4** | **S5** | **S6** | **S7** | **S8** |
| **Maintain local-state** | **F4** | 0 | 0 | 0 | 1 | 0 | 0 | 0 | 0 |
| **Update observers** | **F5** | 0 | 0 | 0 | 1 | 1 | 0 | 0 | 0 |
| **Module= Observer role** | **F4 + F5** | 0 | 0 | 0 | 2 | 1 | 0 | 0 | 0 |

**Fig. A2. *Addition of two functionals.*** Addition of the "Maintain local-state" and "Update observers" functionals gives the row vector containing the overall structors in the *Observer role* in the Observer design Pattern. The structor "Observer" provides only one functional, viz. "Update Observers". The structor "Concrete Observer" provides two different functionals: the "Maintain local-state" functional and the "Update observers" functional. The respective sum is seen in the bottom row.

(NOTE: All Web sites in the above references, last accessed October 2015.)